\documentstyle[12pt]{article}
\begin{document}
\renewcommand{\thefootnote}{\fnsymbol{footnote}}
\def \nnu {$\epsilon_{\mu \nu \rho \sigma} k_1^\nu k_2^\rho
k_3^\sigma $ }
{\hfill HU-TFT-92-50}

\vspace*{0.5cm}

{\bf \hskip 1cm POLARIZATION AND HANDEDNESS IN $\tau\to a_1\nu\to 3\pi\nu$

\hskip 1cm DECAY PRODUCED IN $e^+e^-\to\tau^+\tau^-~$\footnote  
{\rm Invited talk to appear in the conference
proceedings (Universal Academy Press, Tokyo)
of {\it The 10th  International Symposium on High Energy
Spin Physics (SPIN92)}, Nagoya, Japan, November 9-14th 1992}}

\vspace*{1.0cm}

\hskip 1cm N.A. T{\"o}rnqvist

\hskip 1cm Research Institute for High Energy Physics,

\hskip 1cm University of Helsinki, Siltavuorenpenger 20C,

\hskip 1cm P.O. Box 9, SF-00140 Helsinki, Finland.

\vspace*{0.5cm}
\noindent {ABSTRACT } 

We discuss the concept of handedness applied to $\tau$ production
in $e^+e^-$ annihilations with  $\tau$ decaying into
$a_1\nu_\tau \to 3\pi\nu_\tau$. The $a_1\to 3\pi$ decay is particularly
interesting since it (together
with $h_1\to 3\pi$) is the lightest hadronic
decay in which the helicity of the initial state can be determined from
the distribution of the final particle momenta. Thus it is  the first
example of a helicity self-analysing strong decay,
as $\Lambda\to N\pi$ is for weak decays.

Apart from providing a new method for determining the $\tau$ polarization from
the parity conserving part in the weak decay, this reaction provides
a very instructive and useful demonstration of the handedness asymmetry
as a measure of parton polarization.
Thus we believe that experience gained from analyses of
handedness in $\tau\to 3\pi\nu_\tau$ decay
will be very useful when looking for the similar asymmetry of jet handedness
which can measure quark or gluon polarization.

\renewcommand{\thefootnote}{\arabic{footnote}}
\setcounter{footnote}{0}
 \bigskip
{\bf 1. Introduction}
\medskip

This talk is based on recent work together with A. Efremov
and L. Mankie\-wicz \cite{EMT2}. Previously we introduced the concept of
jet handedness~\cite{EMT} (see also Ref.~\cite{NACHT} - 
\cite{COLLINS}), which
will hopefully provide a measurable asymmetry for determining experimentally
the polarization of the initial quark or gluon. There, parity
conservation of strong interactions requires that one has at least three
particles (spinless or spin-averaged) in the final state in order to have
a correlation in the decay distribution with the initial helicity.
Namely, from three particle momenta one can construct a pseudovector
$n_\mu =$~\nnu
which gives, when contracted with the initial polarization pseudovector,
a scalar component in the strong process. Thus the average of the
vector $n_\mu$ can give information
on the initial polarization, provided the
correlation (or analysing power $\alpha$)
is large enough to be measurable.
We called this quantity the handedness vector ($\vec H = \frac 3 2
\alpha \langle \vec n \rangle$).

Because of the complicated process of
jet fragmentation much of the correlation
(i.e. the analysing power $\alpha$)
between polarization and jet handedness is expected to be washed
away when averaging over phase space and/or summing over
different choices of the final three
momenta $k_i$. Therefore, we find it useful and instructive
to look at the simpler situation of
particle decay, where one can even
calculate  the expected handedness and
analysing power $\alpha$.

Looking at the lightest hadronic
decays the process $a_1\to 3\pi$ (together with $h_1\to 3\pi$)
is unique in the sense that it is the lightest
strong decay where one can measure
a handedness, which correlates with
the helicity of the initial state.
Thus this strong decay is similar to the well-known weak decay
$\Lambda \to \pi N$ in the sense that it is "self-analysing"
as to polarization.

One might first think that the $3\pi$
decay of a vector meson ($\omega,\
\phi,\ J/\psi$) could also be self-analysing,
since we do have three final-state particles. However,
these decays do not fulfil a second requirement:
that we must have at least two amplitudes which depend differently
on the initial polarization, since it is the interference between the
two amplitudes which gives the helicity-dependent term. For $\omega\to
3\pi$ we have only one amplitude $\propto \epsilon_{\mu \nu \rho \sigma}
\epsilon^\mu_\omega k_1^\nu k_2^\rho k_3^\sigma $, and therefore we can
only measure the symmetric (tensor) part of the spin-density matrix
($\rho_{ij}\propto$ Re$\varepsilon_i\varepsilon_j^*$), where
$\varepsilon_i$ is the polarization vector of the helicity $i$ state.
Note that it is the antisymmetric part of $\rho_{ij} \propto$
Im$\varepsilon_i\varepsilon_j^*$ which is sensitive to the helicity or
circular polarization vector
${\cal P}_\mu = i \epsilon_{\mu \nu \rho \sigma}
p_\nu \varepsilon_\rho \varepsilon_\sigma^* /m $. The latter is
the true
Pauli-Lubanski pseudovector for angular momentum of spin 1.

We shall consider $a_1$  produced in $e^+e^-$ annihilations
to $\tau^+\tau^-$ with $\tau\to a_1\nu_\tau$. Since the $\tau$ is polarized
(on the average $\approx 14\%$) at the $Z^0$, owing
to the weak interaction,
this polarization is measurable through the one it induces on the
$a_1$. Therefore, a detailed study of the polarization asymmetries in
the
reaction $\tau \to a_1\nu_\tau \to 3\pi\nu_\tau$ is of experimental
interest. However, the $a_1$ is
polarized also by the parity violation in the
decay $\tau\to a_1\nu_\tau$, even if the $\tau$ is unpolarized. This was
studied by K\"uhn and Wagner~\cite{KUHN} and by Feindt~\cite{FEIN}, and
was used by the ARGUS
group~\cite{ARGUS} to measure the helicity of the $\nu_\tau$  to be
left-handed, as expected in the standard model for a sequential
third-generation $\tau$ neutrino.

A unique property of the $\tau \to 3\pi\nu_\tau$
decay is the possibility to separate measurements of the effects due to
the $\tau$ polarization, i.e. the effects of parity
violation in the neutral current, from the effects of left-handedness
of
the $\nu_\tau$, i.e., the parity violation in the charged current. The
latter have to be assumed in studies of $\tau$ polarization with other
decay channels~\cite{ALEPH}--\cite{L3}.
\bigskip

{\bf 2. General analysis}
\medskip

Let us consider the matrix element for $\tau^-$ production in
$e^+\,e^-$ collisions at $Z^0$ energy, followed by its decay into
$\nu_\tau$ and $\pi^+\pi^-\pi^-$ through the $a_1$ intermediate state.
This naturally factorizes into three processes, such that the
first $e^+e^-\to \tau^+\tau^-$ is given by the
cross section $\sigma (e^+e^-\to\tau^+\tau^-)$
of unpolarized $\tau$ and by the
$\tau$ polarization $P_\tau$ along some spin quantization axis $S^\mu$.

Then, the $\tau\to a_1\nu_\tau$ decay can be represented by two
lepton tensors, one  $\bar W^{\mu\nu}(\tau\to a_1\nu_\tau )$
for unpolarized $\tau$'s and another
$W^{\sigma ,\mu\nu}(\tau\to a_1\nu_\tau )$ which will multiply $P_\tau$.
Finally the decay $a_1\to 3\pi$ is determined by a hadronic tensor
$H^{\mu\nu}(a_1\to 3\pi)$.
Thus the cross section for the whole process can be written in
an intuitive form as proportional to
\begin{equation}
\sigma \propto 
{\sigma (e^+e^-\to\tau^+\tau^-)}
 \left[ W^{\mu\nu}(\tau\to a_1\nu_\tau ) +
P_\tau\, W^{\sigma,\mu\nu}(\tau\to a_1\nu_\tau ) \right]
H_{\mu\nu}(a_1\to 3\pi)\ .
\label{e2}
\end{equation}
Now it is important to realize that $W,\ W^\sigma$ and $H$ all
have both a symmetric and an antisymmetric part. Denoting the symmetric
part with a bar and the antisymmetric part with a hat we thus have
\[ W^{\mu\nu} = {\bar W}^{\mu\nu} + {\hat W}^{\mu\nu},\ \
W^{\sigma,\mu\nu} = {\bar W}^{\sigma,\mu\nu} +
{\hat W}^{\sigma,\mu\nu}, \ {\rm and }\ \
H^{\mu\nu} = {\bar H}^{\mu\nu} + {\hat H}^{\mu\nu} \ .
\]

Writing out the explicit expressions for these tensors~\cite{EMT2}
one sees that the symmetric $\bar W^{\mu\nu}$ is parity conserving
($\propto G^2_A+G^2_V$),
and the antisymmetric $\hat W^{\mu\nu}$ is parity violating
($\propto 2i G_AG_V$),
while the symmetric $\bar W^{\sigma ,\mu\nu}$ is parity violating
and the antisymmetric  $\hat W^{\sigma ,\mu\nu}$ is parity conserving.
 Finally the hadronic tensor $H^{\mu\nu}$ is of course
always parity conserving, but contains both a symmetric and an antisymmetric
part.

The $\tau$ polarization  must be measured from
$W^{\sigma,\mu\nu}$, as is  obvious from Eq.~(\ref{e2}).
 Usually one does this from the symmetric and
parity violating part ($\bar W^{\sigma ,\mu\nu})$ whereby one contracts
whith $\bar H^{\mu\nu}$. Then a two-body hadronic decay,
in this case $a_1\to\rho\pi$
(i.e. one can integrate over the $3\pi$ Dalitz plot),
 is sufficient to measure $P_\tau$.

For our purpose it is, however, the parity conserving part  ($\hat
W^{\sigma ,\mu\nu}$) of the $\tau$ decay which is of interest. In order
to pick out this term one must contract with an antisymmetric hadronic
tensor $\hat H^{\mu\nu}$. Then, we can measure the $\tau$ polarization
in a new way, and find a term which can
be present also in a purely strong process such as quark jet fragmentation.

Explicitely these antisymmetric terms have the forms
\begin{eqnarray}
{\hat W}_D^{\sigma,\mu\nu} &=& i (G_A^2 + G_V^2) m_\tau
\epsilon^{\mu\nu\alpha\beta} q_\alpha S_{\beta}\ ,
\label{defWDS} \\
{\hat H}^{\mu\nu}&=& (BW) \left( K_{(2)}^\mu K_{(1)}^\nu - K_{(2)}^\nu
K_{(1)}^\mu
\right) {\mbox Im} (\rho_1^\ast \rho_2) \ ,
\label{defHa}
\end{eqnarray}
\noindent
where  $q$ is the neutrino momentum, BW represents the $a_1$ Breit-Wigner
function (whose form is not important here), $(\rho_1^*\rho_2)$ is the
interference term between the two
crossing $\rho$ bands in the Dalitz plot. Note that now one cannot integrate
over the whole Dalitz plot since then the term from eq.(\ref{defHa})
 vanishes because of  the antisymmetry!
Finally, the hadronic current
\begin{equation}
K_i^\mu = \left( g^{\mu\alpha} - {{k^\mu k^\alpha}\over k^2} \right)
T^i_{\alpha\beta} (k_+ - k_i)^\beta F_{a_1}(k^2)\ ,
\label{defK}
\end{equation}
depends  only on the pion momenta, $k_1,\ k_2, \ k_+$ for the two
negative pions and the positive pion respectively.
The tensor $T^i_{\alpha\beta}$
 contains, in principle, two forms of $a_1\rho\pi$ coupling ($S$- and
$D$-wave),
but experimentally we know that it is almost 100\% $S-$wave.
Finally the $W\to a_1$
formfactor $F_{a_1}$ is irrelevant in our application here, but included
in eq.~(\ref{defK}) for completeness.

After contraction of the two antisymmetric terms one finds a term
\begin{eqnarray}
{\hat M}_1 =\hat W^{\sigma ,\mu\nu} \hat H^{\mu\nu}
&=& 2 \epsilon_{\mu\nu\alpha\beta}
K_{(2)}^\mu K_{(1)}^\nu S^\alpha q^\beta {\mbox Im} \rho_1^\ast \rho_2 \\
&\Rightarrow & 6 n^z \vert{\vec p}\vert {m_\tau
\over {m_T}} (p^0 - \vert{\vec
p}\vert \cos^2{\Psi}) {\mbox Im} \rho_1^\ast \rho_2 \ ,
\end{eqnarray}
\noindent where the second expression is obtained after the averaging
 over the unknown neutrino direction and $\Psi$ is the $\tau$ polar angle
in the $a_1$ CM frame. The relative magnitude of this term to that
of the spin averaged term gives us an asymmetry which is our
handedness
\begin{eqnarray}
{\cal H} &=& {{N(n^z(s_1-s_2)>0) - N(n^z(s_1-s_2)<0)}
\over{N(n^z(s_1-s_2)>0) + N(n^z(s_1-s_2)<0)}} \label{HAND} \\
&=&{{\int d\Omega (BW) \vert {\hat M}_1 \vert} \over
{\int d\Omega (BW) {\bar M}_1}}\, P_\tau (\Theta_a)
 = \bar \alpha P_\tau (\Theta_a)\ ,\label{defCH}
\end{eqnarray}
\noindent where the first equation (\ref{HAND})
is given by the the experimentally
seen number of events of opposite longitudinal handedness, defined
by the sign of $n^z(s_1-s_2)$. Here
$n^z = {\vec e}_a \cdot ({\vec k}_1 \times {\vec k}_2)$ denotes the
third component of the normal to the decay plane when $\vec e_a$ is the
direction of the line of flight of the $a_1$ in the $a_1$ CM, and $s_1,\
s_2$ are the Dalitz plot invariants for the two $\rho$'s.
The second equation (\ref{defCH})  gives  the corresponding theoretical
expression which also defines the analysing
power $\bar \alpha$ which now is a number
which can be calculated theoretically, for the same cuts in phase space
as the data, using the conventional $a_1\to 3\pi$ decay model described
above. At least in one point of phase space ($s_1=m_\rho+m_\rho\Gamma_\rho ,\
s_2=m_\rho-m_\rho\Gamma_\rho$) we know $\alpha$ is nearly
100\% . Thus the average $\bar \alpha$ cannot be too small and by clever cuts
one can increase the analysing power.
\bigskip

{\bf 3. Concluding remarks.} \medskip

The most important result of this work is that one can measure the
polarization of a particle in a parity conserving process from a
three-body final state, and that the $\tau\to 3\pi\nu_\tau$ decay
provides a nice example. Our analysis thus provides a new way to
measure the $\tau$ polarization, and is a very instructive example
for how to proceed in the more complicated process of quark jet
fragmentation. The quark fragmentation
analogy would be that part of the time the quark
fragments into an $a_1$. The incoming quark replaces the $\tau$ of our example
(which is so short lived that it leaves no track, i.e.,
the $\tau$ is  almost as "invisible" as a quark).
Three particles in the jet are assumed to be pions coming from the $a_1$
(or an $a_1$-like object) and the remaining particles of  the jet
replace the unseen neutrino. If the polarized quark ends up in the $a_1$,
the $a_1$ should be polarized and, then the $a_1\to 3\pi$ decay should reveal
the quark polarization in the same way as in the $\tau$ decay example
discussed above. Because of background and combinatorical problems it
is difficult to calculate the analysing power, but provided it is big
enough it should be measurable. Once it is determined for jets
defined in a definite way,
it could be used in other contexts for similar jets to determine the
quark polarization from the jet handedness.

\end{document}